# Acoustic Phonon Spectrum and Thermal Transport in Nanoporous Alumina Arrays

Fariborz Kargar[1,2], Sylvester Ramirez[1,2], Bishwajit Debnath[2,3], Hoda Malekpour[1] , Roger Lake[2,3] and Alexander A. Balandin[1,2]

[1]Phonon Optimized Engineered Materials (POEM) Center, Bourns College of Engineering, University of California – Riverside, Riverside, California 92521 USA

[2]Spins and Heat in Nanoscale Electronic Systems (SHINES) Center, University of California- Riverside, Riverside, California 92521 USA

[3]Laboratory for Terascale and Terahertz Electronics (LATTE), Department of Electrical and Computer Engineering, University of California – Riverside, Riverside, California 92521 USA

**Abstract**

We report results of a combined investigation of thermal conductivity and acoustic phonon spectra in nanoporous alumina membranes with the pore diameter decreasing from $D$=180 nm to 25 nm. The samples with the hexagonally arranged pores were selected to have the same porosity $\phi \approx 13\%$. The Brillouin-Mandelstam spectroscopy measurements revealed bulk-like phonon spectrum in the samples with $D$=180-nm pores and spectral features, which were attributed to spatial confinement, in the samples with 25-nm and 40-nm pores. The velocity of the longitudinal acoustic phonons was reduced in the samples with smaller pores. Analysis of the experimental data and calculated phonon dispersion suggests that both phonon-boundary scattering and phonon spatial confinement affect heat conduction in membranes with the feature sizes $D$<40 nm.

An ability for tuning heat fluxes at nanometer scale offers tremendous benefits for heat removal from the state-of-the-art integrated circuits (ICs) and for increasing efficiency of the thermoelectric (TE) energy conversion. There is growing realization among researchers that effective control of thermal conduction in nanostructures should rely not only on the phonon – boundary scattering but also on the spatially induced modification of the acoustic phonon spectrum [1–10]. The theory predicted that spatial confinement of the acoustic phonons – the main heat carriers in electrical insulators and semiconductors – changes the phonon scattering rates leading to the corresponding modification of the thermal conductivity[1,11,12]. Recent experiments demonstrated that the decrease of the thermal conductivity in nanowires, at least in some cases, cannot be accounted for by the phonon – boundary scattering under an assumption of the bulk acoustic phonon dispersion[3]. The conclusion in Ref. [3] was that the acoustic phonon spectrum has to be modified by the boundaries in order to lead to a strong suppression of the phonon heat conduction[3]. From the other side, there is a growing number of reports showing modification of the acoustic phonon spectrum in nanostructures measured using the Brillouin-Mandelstam light scattering spectroscopy (BMS)[13–20]. However, there have been no reported attempts to correlate the thermal conductivity with the acoustic phonon spectrum in the same structures with the nanometer feature size.

The changes in the acoustic phonon dispersion can be induced either via the stationary boundary conditions in the individual nanostructures, e.g. free-standing nanowires or suspended thin films, or via the periodic boundary conditions, e.g. superlattices. The length scale at which acoustic phonon dispersion should undergo modification and reveal itself at room temperature (RT) is a subject of debates. The traditional metric related to the phonon mean free path (MFP), $\Lambda$, determined from the expression $\Lambda = 3K/C_p \upsilon$ (here $K$ is the phonon thermal conductivity, $C_p$ is the specific heat and $\upsilon$ is the phonon group velocity) has been questioned. The reason is that $\Lambda$ depends strongly on the phonon wavelength, $\lambda$, and the long-wavelength phonons can carry a much larger fraction of heat than it was previously believed[21,22]. These considerations make the correlated study of the thermal conductivity and acoustic phonon spectrum particularly important.

As example structures, we investigated nanoporous alumina films with the hexagonally arranged pores with the average diameter $D$=180 nm, 40 nm and 25 nm. The corresponding inter-pore distances were $H$= 480 nm, 105 nm and 65 nm. The samples produced via a standard electrochemical technique[23] (referred below as A, B, and C) were carefully selected to have the same porosity $\phi = \pi D^2/(2\sqrt{3}H^2)$. For this set specifically, the porosity value was $\phi \approx 13\%$. The constant porosity was important because in the effective medium approximation (EMA), composite samples of the same porosity should have the same thermal conductivity[24–26]. The thickness of alumina films was in the range of 100 μm. Figure 1 shows the scanning electron microscopy (SEM) image of representative samples. One can see a nearly perfect periodic arrangement of the pores.

[Figure 1: Samples and Notations]

The measurements of the cross-plane (perpendicular to the film surface) and in-plane (along the film surface) thermal diffusivity, $\alpha$, were carried out using the transient "laser flash" method (Netzsch LFA). In this technique, the bottom of the sample was illuminated by a flash of a xenon lamp. The temperature of the opposite surface of the sample was monitored with a cryogenically cooled detector. The temperature rise as a function of time was used to extract $\alpha$. The in-plane measurements required a special sample holder described by us elsewhere[27]. The cross-plane thermal conductivity, $K_C$, and in-plane thermal conductivity, $K_I$, were determined from the equation $K_{C,I}=\rho\alpha_{C,I}C_p$, where $\rho$ is the mass density and $C_p$ is the specific heat[28]. The results for $K_C$ are shown in Figure 2. The measured RT values of $K_I$ are summarized in Table I. No clear trend with temperature was observed for $K_I$.

[Figure 2: Experimental Cross-Plane Thermal Conductivity]

**Table I:** Experimental Thermal Conductivity at Room Temperature

| Sample | A: $D$=180 nm | B: $D$=40 nm | C: $D$=25 nm |
|---|---|---|---|
| In-Plane K (W/mK) | 2.5±0.13 | 3.3±0.17 | 2.6±0.13 |
| Cross-Plane K (W/mK) | 1.1±0.05 | 1.0±0.05 | 0.9±0.05 |

The first interesting observation from the experimental data is that the $K_C$ and $K_I$ for $D$=25-nm and $D$=180-nm samples are different despite the same porosity $\phi \approx 13\%$. The second is that both $K_C$ and $K_I$ are much lower than predictions of any of the EMA models. The measured data are consistent with prior study of porous alumina that also reported $K$ in the 0.53 W/mK - 1.62 W/mK range[29]. Our calculation of $K$ of porous alumina using different EMA models gave values between 12 W/mK and 30 W/mK[24–26] for the reported $K$ of bulk alumina without pores, which varies from 15 to 34 Wm$^{-1}$K$^{-1}$ [30]. Given the small Grashof number for alumina and the considered temperature range, the convection and radiation contributions to the heat transport can be neglected[30]. The more than an order-of-magnitude deviation of the experimental $K$ from EMA predictions suggests significant phonon – boundary scattering or phonon spectrum modification.

While the concept of phonons cannot be fully extended to amorphous materials one can still talk about elastic waves and scattering of acoustic phonons in such materials[31,32]. The trend of increasing $K_C$ with $D$ can be rationalized by estimating the phonon – boundary scattering rate $1/\tau_B$. The heat propagates along the pores via the barrier-neck regions with the characteristic lateral dimension of $H-D$ (see Figure 1). Due to periodicity of the structure all barrier-neck regions have the same dimensions. The strength of the acoustic phonon scattering can be roughly estimated from the expression[33] $1/\tau_B = \varsigma(\upsilon/(H-D))[(1-p)/(1+p)]$, where $\varsigma$ is the geometrical factor related to the cross-section shape and $p$ is the specularity parameter determined by the surface roughness. For samples with similar roughness, the ratio of the phonon – boundary scattering rate in the A and C samples, will be $\gamma = (1/\tau_B)_A/(1/\tau_B)_C \approx (\upsilon_A/\upsilon_C)((H_C - D_C)/(H_A - D_A))$. Disregarding the difference in the phonon velocity we obtain $\gamma \approx 0.13$. Optical microscopy and SEM inspection indicated that the $D$=180-nm sample had higher roughness. Assuming conservatively, $p \approx 0$ (diffuse scattering) for A and $p \approx 0.5$, for C, we obtain $\gamma \approx 0.39$. The latter means that the phonon – boundary scattering rate in the sample with larger pores is still smaller despite rougher surface, which translates to higher $K_C$. This conclusion is in line with the study of the thermal conductivity reduction in silicon membranes with hexagonally patterned nearly periodic holes with feature sizes ~ 55 nm[34].

In order to understand whether phonon spectrum modifications can possibly affect the heat conduction in these nanoporous samples we carried out the Brillouin-Mandelstam light scattering spectroscopy (BMS). The experimental setup was based on the six-pass tandem Fabry-Perot interferometer[35–37]. All experiments were carried out using the *p*-unpolarized backscattering mode and different laser incident angles, $\alpha$, ranging from 0 to 60º. The spectra were excited with the solid-state diode-pumped laser operating at $\lambda$=532 nm. The laser light was focused onto a sample by a lens with the numerical aperture of 1.4. The power on the sample was about 100 mW. The scattered light was collected with the same lens and directed to the Fabry-Perot interferometer. Figure 3 shows BMS spectra for the nanoporous samples A, B, and C.

[Figure 3: BMS Spectra]

Owing to the semi-transparent nature of our samples one can expect to observe light scattering from the bulk (i.e. volume) of the material via the elasto-optic scattering mechanism and from the surface of the film via the "ripple" scattering mechanism[37,38]. The interpretation of BMS peaks originating from the elasto-optic scattering requires an accurate knowledge of the refractive index, $n$, of the material. The refractive index was measured by the "prism-coupling" method (Metricon 2010/M)[39]. The experimental values were $n$=1.58 for the cross-plane direction (parallel to the pores) and $n$=1.55 for in-plane direction (perpendicular to the pores) at $\lambda$=532 nm. The Maxwell-Garnett approximation gave a rather close result $n = ((1-\phi)n_m^2 + \phi n_o^2)^{1/2} \approx 1.54$ where $n_m$ is the refractive index of bulk alumina without pores and $n_o$ is the refractive index of air. For the elasto-optic scattering the interacting phonon wave vector, $q$, is given as $q = 4\pi n / \lambda$, which translates to $q$=0.0373 nm$^{-1}$ in our case.

One can see from Figure 3 (a-c) that the acoustic phonon spectrum of 180-nm sample is bulk-like. The peak at ~45 GHz was identified as the longitudinal acoustic (LA) phonon. The measured peak energy corresponds to the phonon velocity $v_L = \omega/q = 2\pi f/q \approx 7580$ m/s (here $f$ is the frequency of the phonon). Near the Brillouin zone (BZ) center the phonon group velocity, which defines thermal transport, coincides with the phase velocity, and it is referred to as phonon velocity in the rest of discussion. The broad shoulder at $f \approx 21$-25 GHz was attributed to the transvers acoustic (TA) phonon. The exact velocity could not have been extracted owing to its large width. BMS spectra for samples with 40-nm and 25-nm pores reveal interesting differences as compared to that of 180-nm sample. The LA phonon peak energy decreases from ~45.0 GHz in 180-nm sample to ~43.4 GHz in the 40-nm sample and ~40.4 GHz in the 25-nm sample. The velocity was extracted for the same incident light angle of θ=40º in order to avoid possible effects of the refractive index anisotropy. The LA phonon velocity in the 25-nm sample is $v_L \approx 6805$ m/s, which corresponds to ~10% reduction. The peaks in 40-nm and 25-nm samples become narrower. However, the peak width cannot be directly related to the phonon life-time in such experiments. The TA peaks in the 40-nm and 25-nm samples become pronounced and show similar reduction in the phonon velocity with decreasing structure feature size from $v_T \approx 3975$ m/s to $v_T \approx 3689$ m/s. To exclude local heating effects from the excitation laser we conducted experiments at different laser powers and accumulation times fixed for all three types of samples. The results were in agreement.

The observed additional peaks at the frequencies below that of TA phonon were attributed to the surface phonons interacting via the "ripple" scattering mechanism[37,38]. In the case of semi-opaque film, the light scattering occurs at or near the surface. In this case, the phonon wave vector selection rules are reduced to the component $q_{//}$ parallel to the surface. This component does not depend on *n* and it is given as $q_\parallel = (4\pi/\lambda)\sin\theta$, where $\theta$ is the incident light angle with respect to the normal to the surface. We established that the surface phonons become visible for θ>20º. The position of these peaks changes with the incident light angle[37,38]. The position of the "volume" LA peak also can change. But the latter is explained not by the dependence of $q_{//}$ on $\theta$ but rather by *n* anisotropy.

For this reason, comparison of $\upsilon_L$ at the same $\theta$ is essential. The effect of the surface phonons on heat conduction (particularly in-plane) requires further studies.

The elasticity theory relates the phonon velocity to material parameters as $\upsilon_L = (E_L / \rho)^{1/2}$ where $E_L$ is the longitudinal Young's modulus and $\rho$ is the mass density. Since all considered samples had the same porosity $\phi \approx 13\%$ one would expect from EMA that should $\upsilon_L$ be the same. However, we observed a trend for decreasing velocity with the decreasing characteristic dimensions of the samples. This suggests that certain modification of the acoustic vibrational spectrum takes place. The elastic continuum theory predicts that the acoustic phonon confinement effects in nanostructures reveal themselves via decreased velocity of the true acoustic modes (defined as those that have $\omega(q=0)=0$) and appearance of the confined quasi-optical subbands ($\omega(q=0)\neq 0$) emanating from the same acoustic polarization branches[11,12,40,41]. The calculated dispersion for our structure in cross-plane direction is shown in Figure 4. The material parameters, extracted from the experimental data for bulk sample (180 nm pores), $\upsilon_T = 7580$ m/s, $\upsilon_T = 3424.9$ m/s and $\rho = 2717.4$ kg/m$^3$ were used in the modeling. One can see from Figure 4 (a) that for a given experimental $q$, the LA peak can consist of a mixture of phonons from several branches. However, the trend of decreasing $\upsilon_L$ and $\upsilon_T$ with reduced feature size is consistent with the experiment (see Figure 4 (b)).

[Figure 4: Calculated spectrum]

Our BMS and computational data indicate that for a given material system the modification of acoustic vibrational spectra became pronounced for the structures with the characteristic dimensions $D=25$ nm ($H$-$D=40$ nm). These dimensions are larger than averaged phonon MFP obtained from the $K$ value. The latter supports recent theoretical considerations of stronger contributions of the long-wavelength phonons with large MFP to heat conduction in solids[21,22]. Noting that in the perturbation theory phonon scattering rate on defects is given as[42] $1/\tau_D \propto \omega^4 / \upsilon^3$ one can conclude that the change in the phonon velocity can noticeably affect $K$.

In summary, we conducted the combined study of thermal conductivity and acoustic phonon spectrum in nanoporous alumina membranes. The analysis of the obtained thermal and phonon spectrum data suggests that both phonon-boundary scattering and phonon confinement affect the heat conduction in the considered alumina nanostructures with the feature sizes $D<40$ nm.

*Acknowledgements*

The work at UC Riverside was supported as part of the Spins and Heat in Nanoscale Electronic Systems (SHINES), an Energy Frontier Research Center funded by the U.S. Department of Energy, Office of Science, Basic Energy Sciences (BES) under Award # SC0012670. The authors thank Prof. C. M. Sotomayor-Torress, Prof. X. Li and Dr. F. Scarponi (JRS Scientific Instruments) for insightful discussions and Prof. V. Fomin and Prof. D.L. Nika for critical reading of the manuscript.

**FIGURE CAPTIONS**

**Figure 1:** Scanning electron microscopy (SEM) image of one of the examined porous alumina samples with notations.

**Figure 2:** Cross-plane thermal conductivity of porous alumina membranes with the pore diameter $D$=25 nm, $D$=40 nm, $D$=180 nm and the corresponding inter-pore distances $H$=65 nm, $H$=105 nm and $H$=480 nm. Despite large difference in feature sizes all samples had the same porosity $\phi$=13%.

**Figure 3:** Brillouin-Mandelstam scattering spectra for the samples with (a) $D$=180 nm, (b) $D$=40 nm and (c) $D$=25 nm. The results are shown for the incident angle $\theta$=40°. Note a shift in the LA and TA phonon peaks with decreasing $D$ and $H$. Multiple peaks in $D$=25-nm sample were attributed to the "ripple" scattering peaks indicated with orange color.

**Figure 4:** (a) Calculated phonon dispersion of a sample with D=25 nm in cross-plane direction. (b) Dispersion of the "true acoustic" branches shown for the samples with D=180 nm ("bulk") and D=25 nm pores. Note the reduced LA and TA phonon velocities are agreement with the experimental data. The experimentally observed LA peak frequency can also be affected by the mode mixing shown in (a). The inset shows displacement distribution for LA mode.

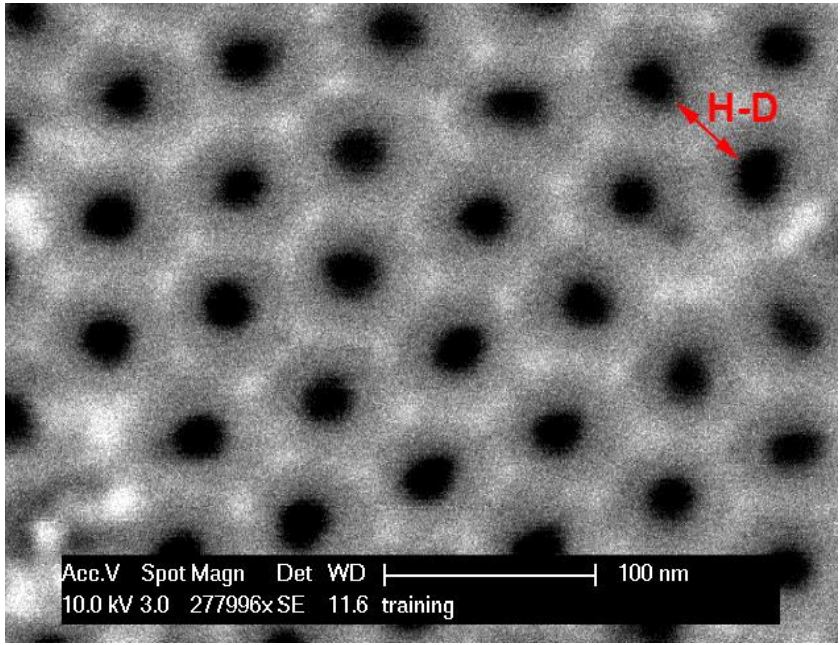

Figure 1

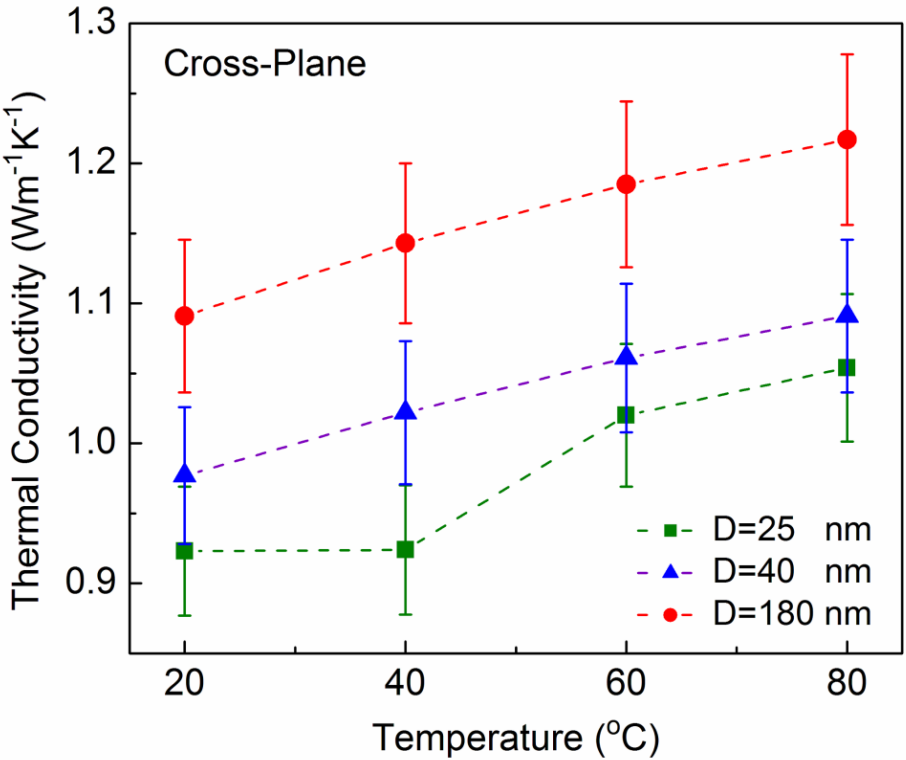

Figure 2

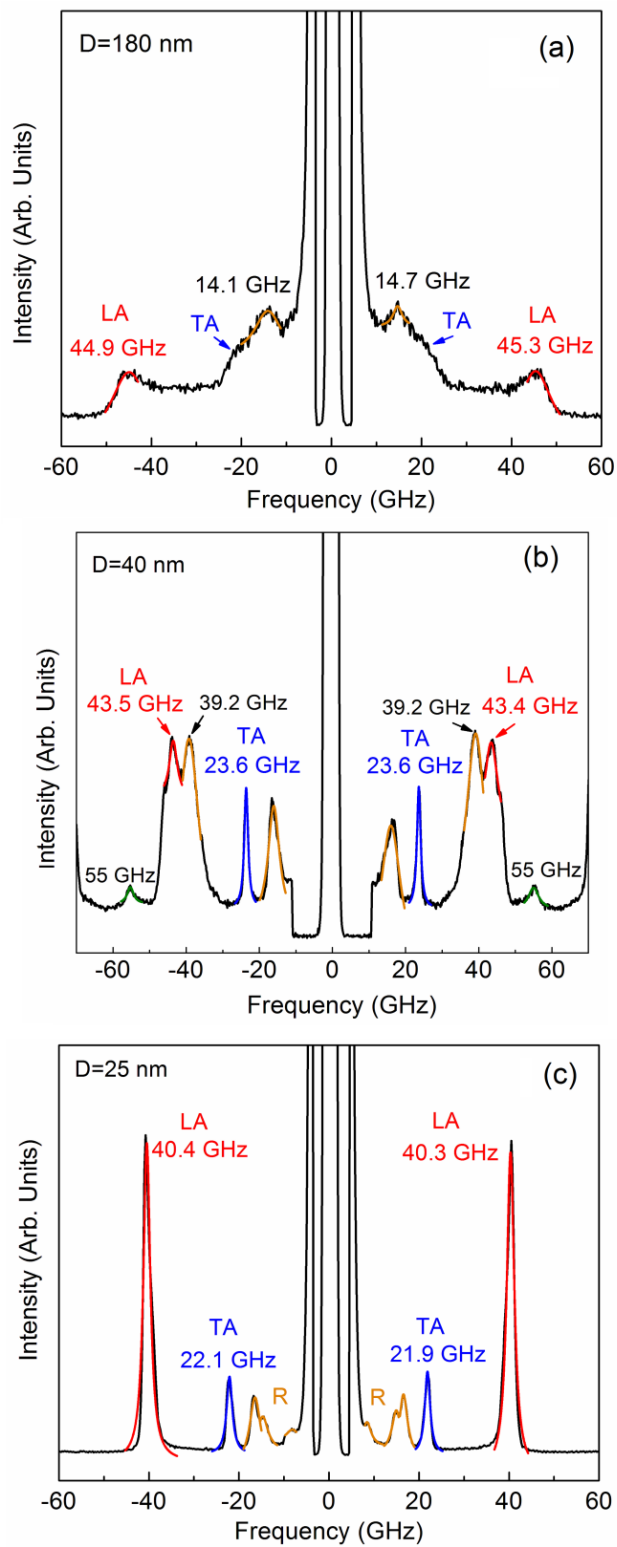

Figure 3

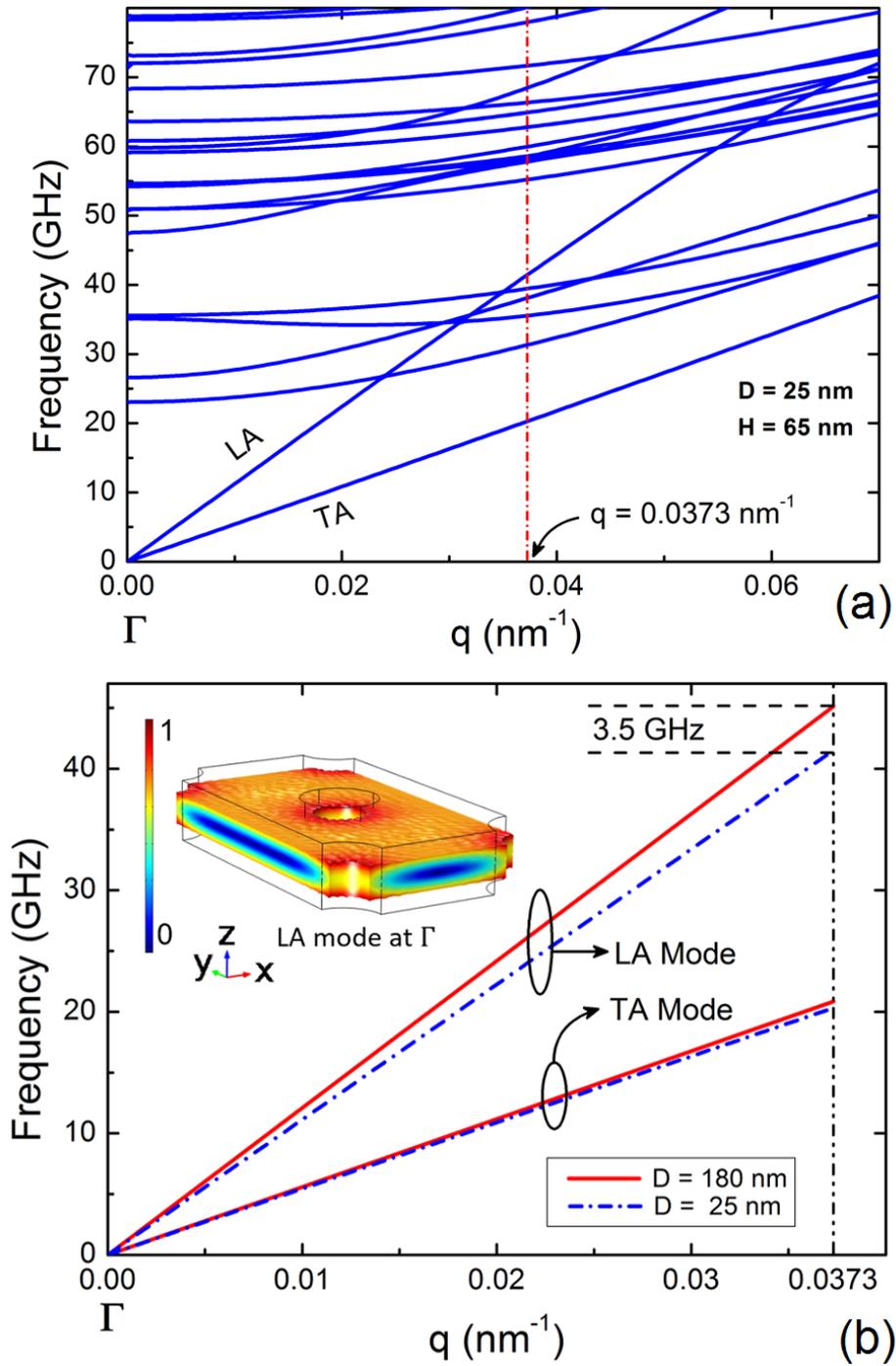

Figure 4